\documentclass[aps,pre,twocolumn,showpacs,superscriptaddress,groupedaddress]{revtex4-2}

\usepackage{graphicx}
\usepackage{dcolumn}
\usepackage{bm}
\usepackage{color} 
\usepackage{amsmath}

\begin{document}
	
\title{Non-standard diffusion under Markovian resetting in bounded domains}
	
	\author{Vicen\c{c} M\'{e}ndez, Axel Mas\'{o}-Puigdellosas and Daniel Campos}
	\affiliation{Grup de F{\'i}sica Estad\'{i}stica.  Departament de F{\'i}sica.
		Facultat de Ci{\`e}ncies. Edifici Cc. Universitat Aut\`{o}noma de Barcelona,
		08193 Bellaterra (Barcelona) Spain}

	\date{\today}
	
	\begin{abstract}
	We consider a walker moving in a one-dimensional interval with absorbing boundaries under the effect of Markovian resettings to the initial position. The walker's motion follows a random walk characterized by a general waiting time distribution between consecutive short jumps. We investigate the existence of an optimal reset rate, which minimizes the mean exit passage time, in terms of the statistical properties of the waiting time probability. Generalizing previous results restricted to Markovian random walks, we here find that, depending on the value of the relative standard deviation of the waiting time probability, resetting can be either (i) never beneficial, (ii) beneficial depending on the distance of the reset to the boundary, or (iii) always beneficial. 
	\end{abstract}

\maketitle

\section{INTRODUCTION}

Brownian motion under restart has been widely studied from a theoretical point of view \cite{EvMa11p}. Depending on the type of random walk and the characteristics of the resetting mechanism, the overall process may reach an equilibrium state \cite{MeCa16} and have an optimal strategy to reach a fixed target \cite{CaMe15}. 
The mean first passage time (MFPT) of a random walker to reach a target located at a given position has been often used to determine the efficiency of resetting for different types of motion \cite{MeMaSaCa21} and reset time distributions \cite{ChSo18,MaCaMe19,BoSo20}. In general, the presence of a safe and fresh reset renders the walker a new opportunity to reach the target whenever it gets far from it. In the particular case where the resetting is Markovian (i.e. it restarts at a constant rate), this process has been shown to exhibit an optimal point for many types of random walk as in the case of a diffusive walker \cite{EvMa11}, subdiffusive \cite{KuGu19,MaMo19}, performing Lévy flights \cite{KuGu15,MaMo21} or under a combination of long-distance jumps interrupted by rests of large duration \cite{MeMaSaCa21}. Random walks under resetting in a bounded domain have also attracted some attention. For example, in Ref. \cite{ChSc15} the authors consider resetting for a diffusing particle moving in a bounded domain with reflecting barriers searching for a target inside the interval and in Ref \cite{PaPeKu19} reactive boundaries have also been considered. Conversely, in \cite{PaPr19,Du19} the exit time of a diffusing particle in a bounded domain with absorbing boundaries is studied when it resets its position to a point $x_0$ at a constant rate $r$. More specifically, they study the passage time from one of the boundaries given that the walker has not reached the other boundary. They find that the condition for which there is an optimal reset rate to exit the region of size $L$ depends on $x_0/L$ only, that is, it is
independent of the diffusion coefficient  (hence, the motion of the particle). 

In this work we generalize the models proposed in \cite{PaPr19,Du19} by considering a random walker whose motion follows a non-standard diffusion under Markovian resetting. Based on the Continuous-Time random walk framework we consider that the walker waits a random time between successive jumps in the diffusive limit, that is, when the characteristic jump length is much lower than $L$. The waiting time is a random variable drawn from a given probability density. We find the criterion for an optimal reset rate which now depends not only on the combination of spatial scales $x_0/L$ but also on the statistical properties of the waiting time density function. We provide numerical studies to support our results.
 
 The paper is organized as follows. In Section \ref{SecCTRWSurv} we introduce the survival probability from a continuous-time random walk perspective and we obtain the survival probability under resetting. In Section \ref{SecMFPT} the mean first exit time is studied and the conditions for which there exists an optimal reset rate are derived. Finally, we present our conclusions in Section \ref{SecConc}.

\section{CTRW AND SURVIVAL PROBABILITY}
\label{SecCTRWSurv}
We consider a walker, initially located at $x_0$, performing a random walk in continuous time within an interval $[0, L]$ in one dimension. The walker can get absorbed by any of these boundaries. In addition, the walker is randomly reset to $x_0$ with a constant rate $r$, i.e., the resetting process is Markovian. We are interested in finding the first-passage time of the walker to see the trade-off between the resetting and the natural absorption of the walker and how it depends on the statistical properties of the waiting time PDF. To see this, we provide an analysis of the survival probability $S_r(x_0, t)$, defined as the probability that the walker has not hit any of the boundaries until time $t$, starting from any $x_0\in [0,L]$. In other words, it estimates
the probability that the walker survives (within the interval) until time $t$. To do this, we first need to solve the Master equation for the random walk and the survival probability in the bounded domain in absence of resetting.

The random walk rules governing the walker motion are as follows. The walker starts from the initial position jumping instantaneously at $t=0$ to a new position where it waits for a time
before proceeding to the next jump. Jump lengths and waiting times
are independent and identically distributed random variables
according to the probability density functions (PDFs) $\Phi(z)$
and $\varphi(t)$, respectively. If $\hat{\varphi}(s)$ and $\tilde{\Phi}(k)$ are their Laplace and Fourier transforms defined by $$\hat{\varphi}(s)=\mathcal{L}_s\left[\varphi(t)\right]=\int_{0}^{\infty}e^{-st}\varphi(t)dt$$
and
$$\tilde{\Phi}(k)=\mathcal{F}_k[\Phi(x)]=\int_{-\infty}^{\infty}e^{ikx}\Phi(x)dx,$$
then the PDF $P(x,t)$ for the walker
position at time $t$ is given by the Montroll-Weiss equation \cite{MoWe65}, which in the Fourier-Laplace space reads 
\begin{equation}
\check{P}(k,s)=\frac{e^{ikx_0}\left[1-\hat{\varphi}(s)\right]}{s\left[1-\hat{\varphi}(s)\tilde{\Phi}(k)\right]}. \label{eq:mw}
\end{equation}
Rearranging
Eq. (\ref{eq:mw}) in the form
\begin{equation}
s\left[\frac{1}{\hat{\varphi}(s)}-\tilde{\Phi}(k)\right]\check{P}(k,s)=e^{ikx_0}\left[\frac{1}{\hat{\varphi}(s)}-1\right],\label{eq:mw2}
\end{equation}
after straightforward algebraic manipulations, we can rewrite the expression as
\begin{equation}
s\check{P}(k,s)-e^{ikx_0}=\hat{K}(s)\left[\tilde{\Phi}(k)-1\right]\check{P}(k,s),\label{eq:mw21}
\end{equation}
where we have introduced the memory kernel
\begin{equation}
\hat{K}(s)=\frac{s\hat{\varphi}(s)}{1-\hat{\varphi}(s)}.\label{eq:mk}
\end{equation}
If we assume that the characteristic jump distance $\sigma$ is very small in comparison with the domain length $L$ ($\sigma \ll L$) then we can consider the continuum limit of $\Phi (x)$ in the Fourier space as $\tilde{\Phi}(k)\simeq1-(\sigma k)^2$. By inverting Eq. (\ref{eq:mw21}) in Fourier-Laplace  we obtain the Master equation for 
non-standard diffusion 
\begin{equation}
\frac{\partial P(x,t)}{\partial t}=\frac{\sigma^2}{2}\int_{0}^{t}K(t-t')\frac{\partial^2 P(x,t')}{\partial x^2}dt'.\label{eq:me}
\end{equation}
Now we are in position to solve \eqref{eq:me} under the boundary and initial conditions
\begin{eqnarray}
    P(x=0,t)&=&P(x=L,t)=0,\nonumber\\
    P(x,t=0)&=&\delta (x-x_0).
    \label{bc}
\end{eqnarray}
Transforming \eqref{eq:me2} by Laplace we obtain 

\begin{equation}
\frac{d^{2}\hat{P}(x,s)}{dx^{2}}-\frac{2s}{\sigma^{2}\hat{K}(s)}\hat{P}(x,s)=-\frac{2\delta(x-x_{0})}{\sigma^{2}\hat{K}(s)}.\label{eq:me2}
\end{equation}
This equation may be solved separately in the two regions I ($0\leq x<x_{0}$)
and II ($x_{0}<x\leq L$). In each region separately the equation
obeys the homogeneous form of Eq. \eqref{eq:me2}, for which the
solutions are

\[
\hat{P}(x,s)=Ae^{\alpha (s) x}+Be^{-\alpha (s) x},
\]
with

\begin{eqnarray}
\alpha (s) \equiv \frac{1}{\sigma}\sqrt{\frac{2s}{\hat{K}(s)}}=\frac{\sqrt{2}}{\sigma}\sqrt{\frac{1}{\hat{\varphi}(s)}-1}.
\label{al}
\end{eqnarray}

In each region we impose the corresponding boundary conditions: $P_{I}(x=0,t)=0$
and $P_{II}(x=L,t)=0.$ In addition, we require that $\hat{P}(x,s)$
is continuous at $x=x_{0}$, i.e., $\hat{P}_{I}(x=x_{0},s)=\hat{P}_{II}(x=x_{0},s)$
and finally, integrating \eqref{eq:me2} from $x_{0}-\varepsilon$ to
$x_{0}+\varepsilon$ and taking the limit $\varepsilon\rightarrow0$
gives
\[
\left(\frac{d\hat{P}_{II}(x,s)}{dx}\right)_{x=x_{0}}-\left(\frac{d\hat{P}_{I}(x,s)}{dx}\right)_{x=x_{0}}=-\frac{2}{\sigma^{2}\hat{K}(s)}.
\]
Considering the above boundary and matching conditions in the solution
for each region one finds after some elementary calculations
\begin{eqnarray}
&\hat{P}&(x,s)=\frac{1}{\sigma}\sqrt{\frac{2}{s\hat{K}(s)}}\frac{1}{\sinh\left(\alpha (s) L\right)}\nonumber\\
&\times&\left\{ \begin{array}{cc}
\sinh\left(\alpha (s)( L-x_0)\right)\sinh\left(\alpha (s) x\right), & 0\leq x\leq x_{0}\\
\sinh\left(\alpha (s)( L-x)\right)\sinh\left(\alpha (s) x_{0}\right), & x_{0}\leq x\leq L
\end{array}\right..
\label{eq:pxs}
\end{eqnarray}

The survival probability in absence of resetting $S_{0}(x_{0},t)$
follows immediately
\begin{eqnarray}
\hat{S}_{0}(x_{0},s)&=&\int_{0}^{L}\hat{P}(x,s)dx\nonumber\\
&=&\frac{1}{s}\left[1-\frac{\cosh\left(\alpha (s)\left( x_{0}-\frac{ L}{2}\right)\right)}{\cosh\left(\alpha (s)\frac{L}{2}\right)}\right].
\label{s0}
\end{eqnarray}
Finally, we can find the survival probability under resetting following
Ref. \cite{ChSo18,MaCaMe19} from the renewal equation
\begin{eqnarray}
&S_r&(x_0,t)=\varphi_{R}^{*}(t)S_0(x_0,t)\nonumber\\
&+&\int_{0}^{t}\varphi_{R}(t')S_0(x,t')S_r(x_0,t-t')dt'
\label{eq:renew0}
\end{eqnarray}
where $\varphi_{R}(t)$ is the PDF of reset times, that is, the probability
that the time elapsed between two consecutive resets is $t,$ and
$\varphi_{R}^{*}(t)=\int_t^{\infty}\varphi_{R}(t')dt'$ is the probability
that no reset has occurred before $t$. The first term on the right-hand side of \eqref{eq:renew0} corresponds to the probability of neither having reached $x_0$, nor a reset has occurred in the period $t\in (0,t]$. The second term is the probability of not having reached $x_0$ when at least one reset has happened at time $t$. In the latter, we account for the probability $S_0(x_0,t)$ of not having reached $x_0$ in the first trip, which finishes when a reset occurs at the random time $t'$, and the probability of not reaching $x_0$ at any other time after the first reset $S_r(x_0,t-t')$.  Applying  the  Laplace transform to \eqref{eq:renew0} we obtain

\begin{eqnarray}
\hat{S}_{r}(x_{0},s)=\frac{\mathcal{L}_{s}\left[\varphi_{R}^{*}(t)S_{0}(x_{0},t)\right]}{1-\mathcal{L}_{s}\left[\varphi_{R}(t)S_{0}(x_{0},t)\right]}
\label{surv}
\end{eqnarray}
Here we will be focusing on the case where resets take place at constant rate $r$, so the corresponding PDF of reset times
follows an exponential distribution 
\begin{eqnarray}
\varphi_{R}(t)=re^{-rt}
\label{expo}
\end{eqnarray}
and \eqref{surv} can be thus rewritten as
\begin{eqnarray}
\hat{S}_{r}(x_{0},s)=\frac{\hat{S}_{0}(x_{0},s+r)}{1-r\hat{S}_{0}(x_{0},s+r)},
\label{surv1}
\end{eqnarray}
with $\hat{S}_0(x_0,s)$ in Eq.\eqref{s0}.

\section{MEAN FIRST EXIT TIME}
\label{SecMFPT}
The time the walker will take to reach for the first any of the boundaries in the presence of resetting at $x_0$ is a random variable distributed, by definition, through the PDF 
\begin{eqnarray}
f_{r}(t)=-\frac{\partial S_r(x_0,t)}{\partial t}.
\end{eqnarray}
The mean first exit time (MFET) is then 
\begin{eqnarray}
T_r(x_0)&=&\int_0^\infty tf_{r}(t)dt=\hat{S}_r(x_0,s = 0)\nonumber\\
&=&\frac{\hat{S}_{0}(x_{0},s=r)}{1-r\hat{S}_{0}(x_{0},s=r)}
\label{tr}
\end{eqnarray}

\subsection{MFET without resetting}
\label{without}

When $r=0$ the survival probability is given by \eqref{s0} and the MFET is, in analogy with \eqref{tr},
\begin{eqnarray}
T_0(x_0)=\hat{S}_0(x_0,s= 0).
\label{t0}
\end{eqnarray}
To compute the above limit we need to know how function $\alpha(s)$ behaves as $s\rightarrow 0$. Taking the derivative of $\alpha(s)$ with respect to $s$
$$
\frac{d\alpha (s)}{ds}=-\frac{1}{\alpha (s) \hat{\varphi}(s)^2 }\frac{d \hat{\varphi}(s) }{ds}>0
$$
since $d\hat{\varphi}(s)/ds<0$. Then $\alpha$ is monotonically increasing with $s$, i.e, 
\begin{eqnarray}
\alpha (s) \rightarrow 0\quad \text{as}\quad s\rightarrow 0.
\label{limit}
\end{eqnarray}
So, for small $s$ we can expand the factor in brackets in \eqref{s0}  to find
\begin{eqnarray}
\hat{S}_0(x_0,s\rightarrow 0)\simeq \frac{\alpha^2(s)}{2s}x_0(L-x_0)+O(\alpha^4 (s)/s).
\label{s0a}
\end{eqnarray}
Likewise, if the waiting time PDF has finite moments then
$$
\frac{1}{\hat{\varphi}(s)}\simeq1+\left\langle t\right\rangle s+s^{2}\left(\left\langle t\right\rangle ^{2}-\frac{\left\langle t^{2}\right\rangle }{2}\right)+O(s^{3}),
$$
and from \eqref{al} one has
\begin{eqnarray}
\alpha^2(s)&=&\frac{2}{\sigma^2}\left(\frac{1}{\hat{\varphi}(s)}-1\right) \nonumber\\
&\simeq& \frac{2}{\sigma^2}\langle t \rangle s+\frac{2s^{2}}{\sigma^2}\left(\left\langle t\right\rangle ^{2}-\frac{\left\langle t^{2}\right\rangle }{2}\right) +O(s^{3})
\label{a2}
\end{eqnarray}
where $\langle t \rangle$ and $\left\langle t^2\right\rangle $ are the mean and the mean square waiting time, respectively. Plugging this result into \eqref{s0a}, the MFET in absence of resetting reads
\begin{eqnarray}
T_0(x_0)=\frac{\langle t \rangle}{\sigma^2}x_0(L-x_0),
\label{a22}
\end{eqnarray}
if the waiting time PDF has finite moments. However, in the case of anomalous diffusion the waiting time PDF has diverging moments and the above result no longer holds. An example is the waiting time PDF
\begin{eqnarray}
\varphi (t)=\frac{t^{\gamma -1}}{\tau^\gamma}E_{\gamma,\gamma}\left(-\frac{t^\gamma}{\tau^\gamma}\right)
\label{ml}
\end{eqnarray}
with $0<\gamma\leq 1$, where
$$E_{\mu,\beta}(z)=\sum_{n=0}^{\infty}\frac{z^{n}}{\Gamma(\mu n+\beta)}$$ is the two parameter Mittag-Leffler function. Its Laplace transform reads
\begin{eqnarray}
\hat{\varphi}(s)=\frac{1}{1+(s\tau)^{\gamma}}.
\label{wta}
\end{eqnarray}
On setting \eqref{wta} into \eqref{al}, the survival probability \eqref{s0a} reads
$$
\hat{S}_0(x_0,s\rightarrow 0)\simeq \frac{\tau x_0(L-x_0)}{\sigma^2(s\tau)^{1-\gamma}}
$$
and inverting by Laplace and taking the time derivative, the first passage PDF is
\begin{eqnarray}
f_0(t)&=&-\frac{\partial S_0(x_0,t)}{\partial t}\nonumber\\
&\simeq& \frac{\gamma x_0(L-x_0)}{\tau\sigma^2\Gamma (1-\gamma)}\left(\frac{\tau}{t}\right)^{1+\gamma}\;\text{as}\;t\rightarrow\infty
\end{eqnarray}
which lacks any finite moments. Note also that from \eqref{t0} and \eqref{s0a} the MFET $T_0(x_0)$ diverges in this case.

\subsection{Optimal reset rate}
We explore here under which conditions there is an optimal rate $r$  minimizing the MFET to reach any of the boundaries.  From \eqref{s0} and \eqref{tr} the MFET in presence of resetting is
\begin{eqnarray}
T_r(x_0)
&=&\frac{1}{r}\left[\frac{\cosh\left(\frac{\alpha (r)L}{2}\right)}{\cosh\left(\alpha (r)\left(x_{0}-\frac{L}{2}\right)\right)}-1\right].
\label{tr2}
\end{eqnarray}
Since the MFET will always increase as $r \rightarrow \infty$, Eq. \eqref{tr2} will necessarily reach a minimum for a specific rate $r$ if the condition
\begin{eqnarray}
\left(\frac{dT_{r}(x_{0})}{dr}\right)_{r=0}<0
\label{cond}
\end{eqnarray}
is fulfilled \cite{PaRe17}. In this case, the value of $r$ which minimizes $T_r(x_0)$ is the optimal reset rate. To understand how this condition is satisfied in terms of the waiting time PDF, we need to analyze separately the cases of waiting time PDFs with both first and second moments finite, only with first moment finite or with both moments diverging.

\subsubsection{Waiting time PDF with finite first and second moments}
\begin{figure}[htbp]
	\includegraphics[width=0.9\hsize]{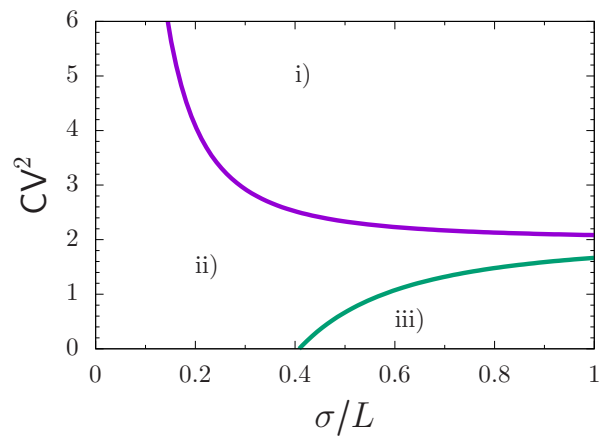}
	\caption{Parameter space diagram for the different existence regions of optimal reset rate}
	\label{phase}
\end{figure}

Taking the limit $r\rightarrow 0$ in \eqref{tr2} and using \eqref{a2} and \eqref{a22} we find
\begin{eqnarray*}
\frac{T_r(x_0)}{T_0(x_0)}&\simeq& 1
+r\left\langle t\right\rangle \left(1-\frac{\left\langle t^{2}\right\rangle }{2\left\langle t\right\rangle ^{2}}-\frac{L^{2}-5x_{0}L+5x_{0}^{2}}{6\sigma^{2}}\right)\\
&+&O(r^2),
\label{tr4}
\end{eqnarray*}
 where the expansion of $T_r(x_0)$ has been normalized dividing by $T_0(x_0)$. Note that this does not affect condition \eqref{cond}, which reads

\begin{eqnarray}
\frac{1-5x_{0}/L+5x_{0}^{2}/L^{2}}{6(\sigma/L)^{2}}>1-\frac{\left\langle t^{2}\right\rangle }{2\left\langle t\right\rangle ^{2}},
\label{cond2}
\end{eqnarray}
holding for any waiting time PDF with finite moments. This condition is a criterion for the existence of an optimal reset rate. When the waiting time PDF is exponential, i.e., the walker's motion follows a standard diffusion, the right hand side of \eqref{cond2} is identically zero and we recover the condition derived in Ref. \cite{PaPr19} (see Eq. (17)). Unlike the case of standard diffusion where the optimal reset rate condition \eqref{cond2} depends on the spatial scales $x_0$ and $L$ only, in general it depends on the  first and second moments of the waiting time PDF. The term of the right hand side of the inequality \eqref{cond2} may be positive, negative or zero. To discuss the possible situations in terms of the statistical properties of the waiting time PDF let us introduce the coefficient of variation or relative standard deviation
$$
\mathsf{CV}=\frac{\sqrt{\left\langle t^{2}\right\rangle }}{\left\langle t\right\rangle }.
$$
Then, from \eqref{cond2} we find the following cases:\\

i) If $\mathsf{CV}^{2}>2+\frac{1}{12}(L/\sigma)^{2}$ condition \eqref{cond2} is satisfied for any value of $x_0/L$. Then an optimal reset rate always exists. In this case, resetting is always beneficial.\\

ii) If $2-\frac{1}{3}(L/\sigma)^{2}<\mathsf{CV}^{2}<2+\frac{1}{12}(L/\sigma)^{2}$ 
the right hand side of condition \eqref{cond2} is positive and an optimal reset rate exists if and only if 
\begin{eqnarray}
\frac{x_{0}}{L}\notin\left(\frac{5-\sqrt{5+\Delta}}{10},\frac{5+\sqrt{5+\Delta}}{10}\right)
\label{cond3}
\end{eqnarray}
with
$$
\Delta=120\left(\frac{\sigma}{L}\right)^{2}\left(1-\frac{\left\langle t^{2}\right\rangle }{2\left\langle t\right\rangle ^{2}}\right)
$$
In this case resetting is beneficial depending on the values of $x_0/L$.\\

iii) If $\mathsf{CV}^{2}<2-\frac{1}{3}(L/\sigma)^{2}$ condition \eqref{cond2} is never satisfied and then no optimal reset rate exists, regardless of the values of $x_0/L$. In this case, resetting is never beneficial. However, this situation is hardly satisfied since we have assumed (see the text below Eq. \eqref{eq:mk}) that the characteristic jump distance is much lower than the domain length, i.e., $\sigma/L\ll 1$. In FIG. \ref{phase} we plot the different cases in a parameter space.
 
To explore the above results we choose a specific waiting time PDF which depends on a parameter $m$. Let us consider the PDF

\begin{eqnarray}
\varphi(t)=\frac{(t/\tau)^m}{\tau\Gamma (m+1)}e^{-t/\tau},\;\; m\geq 0
\end{eqnarray}
which has all moments finite. This includes the exponential distribution by taking $m=0$. The coefficient of variation reads in this case
$$\mathsf{CV}^2=\frac{m+2}{m+1}.$$ For this waiting time PDF the existence condition for optimal reset rate \eqref{cond2} corresponds to the case ii), regardless of $m$. 

Another interesting case is the Pareto PDF
\[
\varphi(t)=\frac{\alpha}{\tau\left(1+\frac{t}{\tau}\right)^{1+\alpha}},\quad\alpha>0.
\]
It has finite moments up to order $n$ if $\alpha>n$, then \eqref{cond2}
holds if $\alpha>2.$ The coefficient of variation is
\[
\mathsf{CV}^{2}=\frac{2\alpha-2}{\alpha-2}
\]
so that an optimal reset rate always exists regardless of the values
of $x_{0}/L$ (case i)) if $2<\alpha<2+24(\sigma/L)^{2}$, which corresponds
to a very narrow case where $\alpha$ is higher than 2 but very close
to 2 provided that $\sigma\ll L.$ The optimal reset rate exists depending
on the values of $x_{0}/L$ (case ii)) if $\alpha>2+24(\sigma/L)^{2}.$ 

\subsubsection{Waiting time PDF with diverging moments}
\begin{figure}[htbp]
	\includegraphics[width=0.8\hsize]{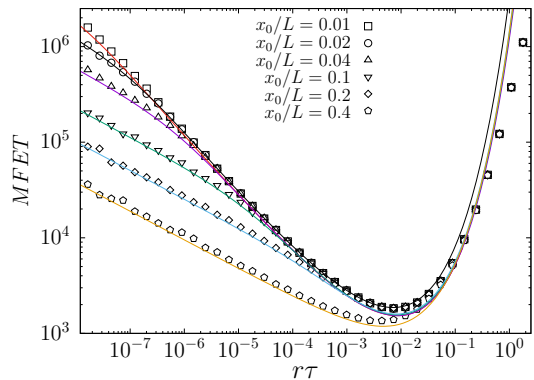}
	\caption{Plot for the MFET vs $r\tau$ for different values of $x_0/L$ when the waiting time PDF is a Mittag-Leffler function with $\gamma=0.7$. Solid curves correspond to theoretical result in Eq. \eqref{tr3} while symbols are the numerical simulations.}
	\label{fig0}
\end{figure}

We explore now the existence of the optimal resetting rate when the walker's motion is subdiffusive, i.e., when the two first moments of waiting time PDF are diverging. This is the case of the Pareto PDF for $0<\alpha<2$ or for the Mittag-Leffler PDF \eqref{ml}. We consider here the latter because it allows us to recover the exponential PDF case for $\gamma =1$.    
Inserting \eqref{wta} in \eqref{eq:mk} and \eqref{tr2} we obtain the MFET
\begin{eqnarray}
T_{r}(x_{0},\gamma)=\frac{1}{r}\left\{\frac{\cosh\left[\frac{\sqrt{2}L}{2\sigma}(r\tau)^{\frac{\gamma}{2}}\right]}{\cosh\left[\frac{\sqrt{2}(x_{0}-L/2)}{\sigma}(r\tau)^{\frac{\gamma}{2}}\right]}-1\right\}.
\label{tr3}
\end{eqnarray}
Since the condition for the existence of an optimal resetting is determined by the behaviour of the MFET near $r=0$, we approximate $T_r(x_0)$ for small $r$ to find
$$
T_{r}(x_{0},\gamma)\simeq\frac{\tau}{\left(r\tau\right)^{1-\gamma}}\frac{x_{0}(L-x_{0})}{\sigma^{2}}
$$
and \eqref{cond} is satisfied regardless of the value of $x_0/L$ because $T_{r}(x_{0},\gamma)$ diverges at $r=0$. In consequence, if the walker moves subdiffusively resetting will always be beneficial. In Fig. \ref{fig0} we plot the MFET showing that there an optimal reset rate for all the values of $x_0/L$, something that can be verified by direct comparison to Monte Carlo simulations (symbols).

To find an analytical expression for the optimal reset rate in terms of the other parameters we need to introduce some approximations. In the continuum spatial limit the characteristic jump size $\sigma$ is very small in comparison with $L$. Considering $L\gg\sigma$ the MFET given in Eq. \eqref{tr3} can be rewritten in the form
\begin{figure}[htbp]
	\includegraphics[width=0.8\hsize]{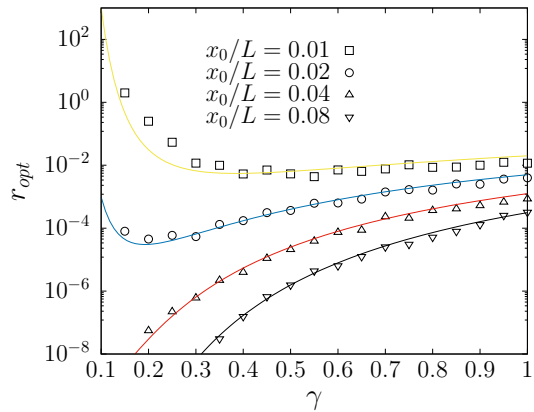}
	\caption{Plot for the optimal reset rate $r_{opt}$ vs. the anomalous exponent $\gamma$ (for $\sigma =1e^{-3}$ and $L=1$),}
	\label{fig1}
\end{figure}
\begin{eqnarray}
T_{r }\sim z^{-2/\gamma}\left(e^{2z\delta}-1\right)
\end{eqnarray}
where $z=(r\tau)^{2/\gamma}/\sqrt{2}$ and we have assumed  $\delta = x_0/\sigma\gg 1$. Taking the derivative of $T_r$ and equating to zero we find that the optimal reset rate $r_{opt}$ is given by
\begin{eqnarray}
r_{opt}\simeq \frac{1}{\tau}\left(\frac{\sqrt{2}\sigma}{\gamma x_0}\right)^{2/\gamma}.
\label{ropt}
\end{eqnarray}

 Note that the approximation $x_0/\sigma\gg 1$ holds in the continuum spatial limit if $x_0 \sim O(L)$. In figure \ref{fig1} we compare \eqref{ropt}  with the numerical simulations; note that due to the spatial symmetry in the domain, taking a value for $x_0$ close to 0, namely $x_0=\epsilon\ll 1$, is equivalent to $x_0=L-\epsilon$. In general the agreement observed in figure \ref{fig1} is excellent but for small values of $\gamma$ the waiting times are eventually very large and this requires very high computational time which reduces the number of realizations and the accuracy. It is remarkable, in particular, the non-monotonic behavior exhibited by $r_{opt}$, and the fact that for $\gamma$ small the optimal reset rate can get modified by several orders of magnitude just by slightly changing the reset position $x_0$. This is a nontrivial consequence of the interplay between the diverging waiting times (which lead to a diverging MFET, as seen in Section \ref{without}) and the timescale for resetting, $r^{-1}$. There is a necessity to include resettings to stop long waiting times, but this must be done without compromising too much the probability to reach the boundary.  
 
Finally, note that inserting \eqref{ropt} into \eqref{tr3} the minimal MFET reads
 \begin{eqnarray}
T_{r }^*\simeq \tau \left(\frac{\gamma x_{0}}{\sqrt{2}\sigma}\right)^{2/\gamma}e^{2/\gamma}\sim\left(\frac{x_{0}}{\sigma}\right)^{2/\gamma}.
\end{eqnarray}

\subsection{Optimal search strategy}
\begin{figure}[htbp]
	\includegraphics[width=0.8\hsize]{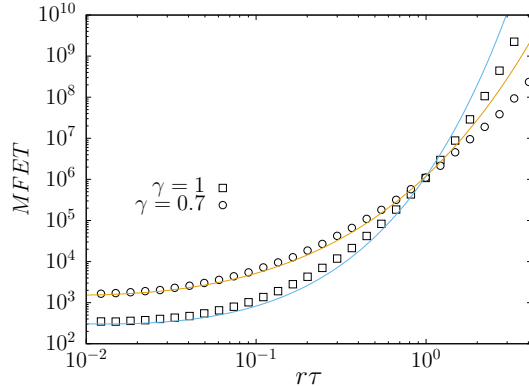}
	\caption{Plot for $T_r(x_0,\gamma)$ vs $r\tau$. Symbols correspond to numerical simulations and solid curves are the theoretical predictions given by Eq.\eqref{tr3}. $L=1$, $x_0=0.1$ and $\sigma=0.01$.   }
	\label{fig3}
\end{figure}

Next we investigate which is the search strategy which has a lower MFET for a given reset rate $r$, that is, what is the waiting time PDF (exponential or Mittag-Leffler) that has a lower MFET for a given $r$. To do this we compare $T_r(x_0,\gamma)$ with $T_r(x_0,\gamma=1)$. 
For fixed $x_0$, $L$ and $\gamma$, the equation $T_r(x_0,\gamma)=T_r(x_0,\gamma=1)$ has a unique solution for $r\tau=1$. Defining $r\tau =1-\epsilon$ with $|\epsilon|\ll 1$, the quotient between both MFETs can be expanded about $r\tau=1$ 
$$
\frac{T_{r}(x_{0},\gamma)}{T_{r}(x_{0},\gamma=1)}\simeq1+\frac{1-\gamma}{\sigma\sqrt{2}}A\epsilon+O(\epsilon^{2})
$$
where
$$
A=a\frac{(L-x_{0})b(a^{2}-1)+x_{0}(b^{2}-a^{2})}{(a-1)(b-a)(a^{2}+b)}
$$
and
$$
a=e^{\sqrt{2}x_{0}/\sigma},\;b=e^{\sqrt{2}L/\sigma}.
$$
Clearly $1<a<b$, so that $A>0$. From this we conclude that if $r\tau<1$ (low reset rates) (i.e., $\epsilon>0$) then
$$
T_{r}(x_{0},\gamma)>T_{r}(x_{0},\gamma=1)
$$
and the search strategy with exponential waiting time is optimal. If $r\tau>1$ (high reset rates) (i.e., $\epsilon<0$) then
$$
T_{r}(x_{0},\gamma)<T_{r}(x_{0},\gamma=1)
$$
and the search strategy with anomalous waiting time is the optimal. This is confirmed by numerical simulations as shown in Figure  \ref{fig3}. While the crossing point is robustly found at $r \tau =1$, note that simulations for $r\tau >1$ are not accurate because in this regime the reset process is too fast compared to the characteristic time that the walker needs to reach the boundary from $x_0$. In consequence, the walker spends all of the time resetting again and again until by chance an extreme value of the resetting time appears and so it opens the possibility for the walker to reach the boundary. As a result of this dynamics, the computing time for simulations increase exponentially with $r$, so that the number of realizations that can be computed in a reasonable time reduces very much, so compromising the accuracy of the results. 

\section{Conclusions}
\label{SecConc}
By generalizing the model studied in \cite{PaPr19} for a diffusive random walk in a finite interval with resets, we have been able to explore interesting, and  previously unreported, regimes of optimal behavior of the MFET as a function of the reset rate. A formal expression for the mean exit time has been obtained for general waiting time (Eq.\eqref{tr2}) and jump length distributions in the limit where the length of the interval $L$ is large in comparison with the jump distance $\sigma$. From that, we find that the existence of an optimal reset rate to exit the interval depends exclusively on the spatial scales only for a very particular choice of the waiting time distribution of the walker (exponential distribution). Conversely, whenever the waiting times are not Markovian, the optimality of resetting depends also on the shapes of both the jump length and the waiting time distribution of the walker. 
Finally, we have also found that the optimal strategy to exit the interval given a reset rate $r$ depends on the rate itself. For low reset rates, walkers with exponential waiting times are found to be optimal and, when resetting is more frequent, anomalous waiting times optimize the process. These results open new questions as whether the casuistic herein found would hold for a more general non-Markovian resetting. Also, a general investigation of a mixed absorbing-reflecting boundaries is still lacking in the resetting literature.

\bibliography{references}

\begin{thebibliography}{18}
\expandafter\ifx\csname natexlab\endcsname\relax\def\natexlab#1{#1}\fi
\expandafter\ifx\csname bibnamefont\endcsname\relax
  \def\bibnamefont#1{#1}\fi
\expandafter\ifx\csname bibfnamefont\endcsname\relax
  \def\bibfnamefont#1{#1}\fi
\expandafter\ifx\csname citenamefont\endcsname\relax
  \def\citenamefont#1{#1}\fi
\expandafter\ifx\csname url\endcsname\relax
  \def\url#1{\texttt{#1}}\fi
\expandafter\ifx\csname urlprefix\endcsname\relax\def\urlprefix{URL }\fi
\providecommand{\bibinfo}[2]{#2}
\providecommand{\eprint}[2][]{\url{#2}}

\bibitem[{\citenamefont{Evans et~al.}(2020)\citenamefont{Evans, Majumdar, and
  Schehr}}]{EvMa11p}
\bibinfo{author}{\bibfnamefont{M.~R.} \bibnamefont{Evans}},
  \bibinfo{author}{\bibfnamefont{S.~N.} \bibnamefont{Majumdar}},
  \bibnamefont{and} \bibinfo{author}{\bibfnamefont{G.}~\bibnamefont{Schehr}},
  \bibinfo{journal}{Journal of Physics A: Mathematical and Theoretical}
  \textbf{\bibinfo{volume}{53}}, \bibinfo{pages}{193001}
  (\bibinfo{year}{2020}),
  \urlprefix\url{https://doi.org/10.1088/1751-8121/ab7cfe}.

\bibitem[{\citenamefont{M\'endez and Campos}(2016)}]{MeCa16}
\bibinfo{author}{\bibfnamefont{V.}~\bibnamefont{M\'endez}} \bibnamefont{and}
  \bibinfo{author}{\bibfnamefont{D.}~\bibnamefont{Campos}},
  \bibinfo{journal}{Phys. Rev. E} \textbf{\bibinfo{volume}{93}},
  \bibinfo{pages}{022106} (\bibinfo{year}{2016}),
  \urlprefix\url{https://link.aps.org/doi/10.1103/PhysRevE.93.022106}.

\bibitem[{\citenamefont{Campos and M\'endez}(2015)}]{CaMe15}
\bibinfo{author}{\bibfnamefont{D.}~\bibnamefont{Campos}} \bibnamefont{and}
  \bibinfo{author}{\bibfnamefont{V.}~\bibnamefont{M\'endez}},
  \bibinfo{journal}{Phys. Rev. E} \textbf{\bibinfo{volume}{92}},
  \bibinfo{pages}{062115} (\bibinfo{year}{2015}),
  \urlprefix\url{https://link.aps.org/doi/10.1103/PhysRevE.92.062115}.

\bibitem[{\citenamefont{M{\'e}ndez et~al.}(2021)\citenamefont{M{\'e}ndez,
  Mas{\'o}-Puigdellosas, Sandev, and Campos}}]{MeMaSaCa21}
\bibinfo{author}{\bibfnamefont{V.}~\bibnamefont{M{\'e}ndez}},
  \bibinfo{author}{\bibfnamefont{A.}~\bibnamefont{Mas{\'o}-Puigdellosas}},
  \bibinfo{author}{\bibfnamefont{T.}~\bibnamefont{Sandev}}, \bibnamefont{and}
  \bibinfo{author}{\bibfnamefont{D.}~\bibnamefont{Campos}},
  \bibinfo{journal}{Phys. Rev. E} \textbf{\bibinfo{volume}{103}},
  \bibinfo{pages}{022103} (\bibinfo{year}{2021}),
  \urlprefix\url{https://link.aps.org/doi/10.1103/PhysRevE.103.022103}.

\bibitem[{\citenamefont{Chechkin and Sokolov}(2018)}]{ChSo18}
\bibinfo{author}{\bibfnamefont{A.}~\bibnamefont{Chechkin}} \bibnamefont{and}
  \bibinfo{author}{\bibfnamefont{I.~M.} \bibnamefont{Sokolov}},
  \bibinfo{journal}{Phys. Rev. Lett.} \textbf{\bibinfo{volume}{121}},
  \bibinfo{pages}{050601} (\bibinfo{year}{2018}),
  \urlprefix\url{https://link.aps.org/doi/10.1103/PhysRevLett.121.050601}.

\bibitem[{\citenamefont{Mas\'o-Puigdellosas
  et~al.}(2019)\citenamefont{Mas\'o-Puigdellosas, Campos, and
  M\'endez}}]{MaCaMe19}
\bibinfo{author}{\bibfnamefont{A.}~\bibnamefont{Mas\'o-Puigdellosas}},
  \bibinfo{author}{\bibfnamefont{D.}~\bibnamefont{Campos}}, \bibnamefont{and}
  \bibinfo{author}{\bibfnamefont{V.}~\bibnamefont{M\'endez}},
  \bibinfo{journal}{Phys. Rev. E} \textbf{\bibinfo{volume}{99}},
  \bibinfo{pages}{012141} (\bibinfo{year}{2019}),
  \urlprefix\url{https://link.aps.org/doi/10.1103/PhysRevE.99.012141}.

\bibitem[{\citenamefont{Bodrova and Sokolov}(2020)}]{BoSo20}
\bibinfo{author}{\bibfnamefont{A.~S.} \bibnamefont{Bodrova}} \bibnamefont{and}
  \bibinfo{author}{\bibfnamefont{I.~M.} \bibnamefont{Sokolov}},
  \bibinfo{journal}{Phys. Rev. E} \textbf{\bibinfo{volume}{101}},
  \bibinfo{pages}{062117} (\bibinfo{year}{2020}),
  \urlprefix\url{https://link.aps.org/doi/10.1103/PhysRevE.101.062117}.

\bibitem[{\citenamefont{Evans and Majumdar}(2011)}]{EvMa11}
\bibinfo{author}{\bibfnamefont{M.~R.} \bibnamefont{Evans}} \bibnamefont{and}
  \bibinfo{author}{\bibfnamefont{S.~N.} \bibnamefont{Majumdar}},
  \bibinfo{journal}{Phys. Rev. Lett.} \textbf{\bibinfo{volume}{106}},
  \bibinfo{pages}{160601} (\bibinfo{year}{2011}),
  \urlprefix\url{https://link.aps.org/doi/10.1103/PhysRevLett.106.160601}.

\bibitem[{\citenamefont{Ku\ifmmode~\acute{s}\else \'{s}\fi{}mierz and
  Gudowska-Nowak}(2019)}]{KuGu19}
\bibinfo{author}{\bibfnamefont{L.}~\bibnamefont{Ku\ifmmode~\acute{s}\else
  \'{s}\fi{}mierz}} \bibnamefont{and}
  \bibinfo{author}{\bibfnamefont{E.}~\bibnamefont{Gudowska-Nowak}},
  \bibinfo{journal}{Phys. Rev. E} \textbf{\bibinfo{volume}{99}},
  \bibinfo{pages}{052116} (\bibinfo{year}{2019}),
  \urlprefix\url{https://link.aps.org/doi/10.1103/PhysRevE.99.052116}.

\bibitem[{\citenamefont{Masoliver and Montero}(2019)}]{MaMo19}
\bibinfo{author}{\bibfnamefont{J.}~\bibnamefont{Masoliver}} \bibnamefont{and}
  \bibinfo{author}{\bibfnamefont{M.}~\bibnamefont{Montero}},
  \bibinfo{journal}{Phys. Rev. E} \textbf{\bibinfo{volume}{100}},
  \bibinfo{pages}{042103} (\bibinfo{year}{2019}),
  \urlprefix\url{https://link.aps.org/doi/10.1103/PhysRevE.100.042103}.

\bibitem[{\citenamefont{Ku\ifmmode~\acute{s}\else \'{s}\fi{}mierz and
  Gudowska-Nowak}(2015)}]{KuGu15}
\bibinfo{author}{\bibfnamefont{L.}~\bibnamefont{Ku\ifmmode~\acute{s}\else
  \'{s}\fi{}mierz}} \bibnamefont{and}
  \bibinfo{author}{\bibfnamefont{E.}~\bibnamefont{Gudowska-Nowak}},
  \bibinfo{journal}{Phys. Rev. E} \textbf{\bibinfo{volume}{92}},
  \bibinfo{pages}{052127} (\bibinfo{year}{2015}),
  \urlprefix\url{https://link.aps.org/doi/10.1103/PhysRevE.92.052127}.

\bibitem[{\citenamefont{Majumdar et~al.}(2021)\citenamefont{Majumdar, Mounaix,
  Sabhapandit, and Schehr}}]{MaMo21}
\bibinfo{author}{\bibfnamefont{S.~N.} \bibnamefont{Majumdar}},
  \bibinfo{author}{\bibfnamefont{P.}~\bibnamefont{Mounaix}},
  \bibinfo{author}{\bibfnamefont{S.}~\bibnamefont{Sabhapandit}},
  \bibnamefont{and} \bibinfo{author}{\bibfnamefont{G.}~\bibnamefont{Schehr}},
  \bibinfo{journal}{Journal of Physics A: Mathematical and Theoretical}
  \textbf{\bibinfo{volume}{55}}, \bibinfo{pages}{034002}
  (\bibinfo{year}{2021}),
  \urlprefix\url{https://doi.org/10.1088/1751-8121/ac3fc1}.

\bibitem[{\citenamefont{Christou and Schadschneider}(2015)}]{ChSc15}
\bibinfo{author}{\bibfnamefont{C.}~\bibnamefont{Christou}} \bibnamefont{and}
  \bibinfo{author}{\bibfnamefont{A.}~\bibnamefont{Schadschneider}},
  \bibinfo{journal}{Journal of Physics A: Mathematical and Theoretical}
  \textbf{\bibinfo{volume}{48}}, \bibinfo{pages}{285003}
  (\bibinfo{year}{2015}),
  \urlprefix\url{https://doi.org/10.1088%2F1751-8113%2F48%2F28%2F285003}.

\bibitem[{\citenamefont{Pal et~al.}(2019)\citenamefont{Pal, Castillo, and
  Kundu}}]{PaPeKu19}
\bibinfo{author}{\bibfnamefont{A.}~\bibnamefont{Pal}},
  \bibinfo{author}{\bibfnamefont{I.~P.} \bibnamefont{Castillo}},
  \bibnamefont{and} \bibinfo{author}{\bibfnamefont{A.}~\bibnamefont{Kundu}},
  \bibinfo{journal}{Phys. Rev. E} \textbf{\bibinfo{volume}{100}},
  \bibinfo{pages}{042128} (\bibinfo{year}{2019}),
  \urlprefix\url{https://link.aps.org/doi/10.1103/PhysRevE.100.042128}.

\bibitem[{\citenamefont{Pal and Prasad}(2019)}]{PaPr19}
\bibinfo{author}{\bibfnamefont{A.}~\bibnamefont{Pal}} \bibnamefont{and}
  \bibinfo{author}{\bibfnamefont{V.~V.} \bibnamefont{Prasad}},
  \bibinfo{journal}{Physical Review E} \textbf{\bibinfo{volume}{99}}
  (\bibinfo{year}{2019}).

\bibitem[{\citenamefont{Durang et~al.}(2019)\citenamefont{Durang, Lee, Lizana,
  and Jeon}}]{Du19}
\bibinfo{author}{\bibfnamefont{X.}~\bibnamefont{Durang}},
  \bibinfo{author}{\bibfnamefont{S.}~\bibnamefont{Lee}},
  \bibinfo{author}{\bibfnamefont{L.}~\bibnamefont{Lizana}}, \bibnamefont{and}
  \bibinfo{author}{\bibfnamefont{J.-H.} \bibnamefont{Jeon}},
  \bibinfo{journal}{Journal of Physics A: Mathematical and Theoretical}
  \textbf{\bibinfo{volume}{52}}, \bibinfo{pages}{224001}
  (\bibinfo{year}{2019}),
  \urlprefix\url{https://doi.org/10.1088%2F1751-8121%2Fab15f5}.

\bibitem[{\citenamefont{Montroll and Weiss}(1965)}]{MoWe65}
\bibinfo{author}{\bibfnamefont{E.~W.} \bibnamefont{Montroll}} \bibnamefont{and}
  \bibinfo{author}{\bibfnamefont{G.~H.} \bibnamefont{Weiss}},
  \bibinfo{journal}{Journal of Mathematical Physics}
  \textbf{\bibinfo{volume}{6}}, \bibinfo{pages}{167} (\bibinfo{year}{1965}),
  \urlprefix\url{https://doi.org/10.1063%2F1.1704269}.

\bibitem[{\citenamefont{Pal and Reuveni}(2017)}]{PaRe17}
\bibinfo{author}{\bibfnamefont{A.}~\bibnamefont{Pal}} \bibnamefont{and}
  \bibinfo{author}{\bibfnamefont{S.}~\bibnamefont{Reuveni}},
  \bibinfo{journal}{Phys. Rev. Lett.} \textbf{\bibinfo{volume}{118}},
  \bibinfo{pages}{030603} (\bibinfo{year}{2017}),
  \urlprefix\url{https://link.aps.org/doi/10.1103/PhysRevLett.118.030603}.

\end{thebibliography}

\end{document}